%% file: rfi.tex
\documentclass[12pt]{article}
\pdfoutput=0

\usepackage{fancyheadings}

\usepackage{ulem}
\usepackage{psfrag}
\usepackage[dvips]{graphicx}

\usepackage{wrapfig}
\usepackage{shadow}
\usepackage{color}
\usepackage{dcolumn}% Align table columns on decimal point

\usepackage{amsmath}
\usepackage{amsfonts}
\usepackage{amssymb}
\usepackage{bm}% bold mathg

\usepackage{sectsty}
\sectionfont{\large}

\usepackage{comment} 
\usepackage{hyperref}
\usepackage{breakurl}

\newcommand{\beq}{\begin{equation}}
\newcommand{\eeq}{\end{equation}}
\newcommand{\HRule}{\rule{\linewidth}{0.5mm}}

\begin{document}

\pagestyle{fancyplain}

\pagenumbering{arabic}

%\medskip

\input{./title.tex}

\section*{\sc Description}

We present a mission concept, the Geostationary Antenna for Disturbance-Free Laser 
Interferometry (GADFLI), for a space-based gravitational-wave interferometer consisting of three satellites in geostationary orbit
around the Earth\footnote{After submission of this mission concept to the NASA-issued Request for Information (RFI), we became aware of a
very similar mission concept also submitted to the RFI \cite{Massimo}. That concept and the one we propose, though strikingly
similar, were nonetheless developed independently and submitted to the RFI simultaneously.}.  Compared to the nominal design of the
Laser Interferometer Space Antenna (LISA), this concept has the advantage of significantly decreased requirements on the telescope size and
laser power, decreased launch mass, 
and substantially improved shot noise resulting from the much shorter 73000 km armlengths ($2\times \cos(30^{\rm o})\times 42164$ km $\approx 73000$ km).
We note that the constellation and the size of the Earth are approximately to scale in the graphic on the title page.
Communications are also simplified in this design, and the possibility of a servicing mission in the event of a single satellite
failure is more viable.  The three satellites are identical, and the cost of a 120 degree phase change for 2 satellites
is minimal, so that the principle launch cost will be the transfer of these three satellites, which are lighter than the three
LISA satellites, to a single geostationary orbit.

The primary disadvantage is the potentially diminished performance at frequencies below a mHz due to increased proof mass
acceleration noise.
A perceived secondary disadvantage is the need for station-keeping, since the Sun and Moon provide  
torques out of the constellation's orbital plane due to its 23 degree inclination to the ecliptic, which would cause
a relative drift among the sciencecrafts.  However, this drift is at the level of $\Delta v=45$ m/s each year, 
and is primarily directed out of the plane of measurement, which implies a Doppler shift well below 45 MHz per year for a micrometer wavelength laser.
Conservatively estimating the Doppler shift at 25 MHz, phasemeter sampling at 50 MHz would therefore allow operation for at least two years
without station keeping.
Such a sampling rate is thought to be consistent with the capabilities
of available phasemeters in the range of cost of the nominal LISA design \cite{Ira}.  We therefore limit the mission lifetime
of this proposal to 2 years, to avoid the need for the additional hardware required for station keeping and its impact on cost.

We consider three scenarios, where the noisier environment in Earth's orbit results in a failure to meet the DRS specifications of 
$3 \times 10^{-15} {\rm m}/{\rm s}^2/\sqrt{\rm Hz}$ and performance is worse by an order of magnitude, where the DRS specification is met despite
the noisier environment, and the very optimistic scenario where the DRS specification is exceeded by an order of magnitude.  Given that none
of these levels of accuracy has been achieved at these frequencies before, there is considerable uncertainty regarding what will be achievable
when the final mission hardware is assembled.  We note that our optimistic choice places the residual DRS noise at the same level
as the galactic foreground of white dwarf binaries, so that further improvement would not benefit the science performance.  This also
motivates our use of the term ``displacement free'' in the concept name, since an order-of-magnitude improvement over the nominal
LISA DRS would render displacement noise a non-factor in this design.

Compared to the OMEGA mission, which is another geocentric constellation with a much longer armlength,
GADFLI again has decreased requirements on telescope size and laser power, lower mass and decreased shot noise, and may have the same disadvantage
of increased acceleration noise, though this is less clear than in the comparison with LISA.  An additional advantage of GADFLI is the 23 degree
inclination of an equatorial geocentric orbit with respect to the ecliptic, which will prevent the GADFLI optics from ever being exposed
to near-direct sunlight.  This will also provide greater thermal stability, as the satellites are never eclipsed by the Earth over the course of their
orbit.  Because of the longer armlength, OMEGA must orbit in the Earth-Moon plane, which is only inclined 5 degrees to the ecliptic, and must
therefore be outfitted with yet-to-be-developed filters, in order to prevent damage to the optics during their exposure to direct sunlight.

\section*{\sc Specifications}

The GADFLI constellation will fly at the $\sim$35800 km elevation above sea level
required for geostationarity, which corresponds to a constellation armlength
of approximately 73000 km.  The mirror size can be decreased to 15 cm without resulting in a significant impact to sensitivity.
Likewise, the laser power can be reduced to 0.7 W.  In Fig.~1, we plot the rms strain sensitivity
for three variations of the GADFLI concept, as well as the nominal LISA design and the recently-selected NGO configuration.
We have followed \cite{Larson} to implement the sensitivity, including the Hils and Bender white dwarf confusion
estimate \cite{Hils}.  We have verified that, apart from the sinusoidal contributions to the response that we have neglected and which
are unimportant for signal-to-noise ratio (SNR) calculations, we duplicate the results found from running the online sensitivity curve generator \cite{Sens}
provided by the lead author of \cite{Larson}.
Although the noise ``bucket'' of the GADFLI curve is less sensitive, the frequency band of GADFLI is more optimal given the expectation of
likely sources, as we will show in the following section.
\begin{center}
\begin{figure}
\includegraphics[width=0.95\textwidth]{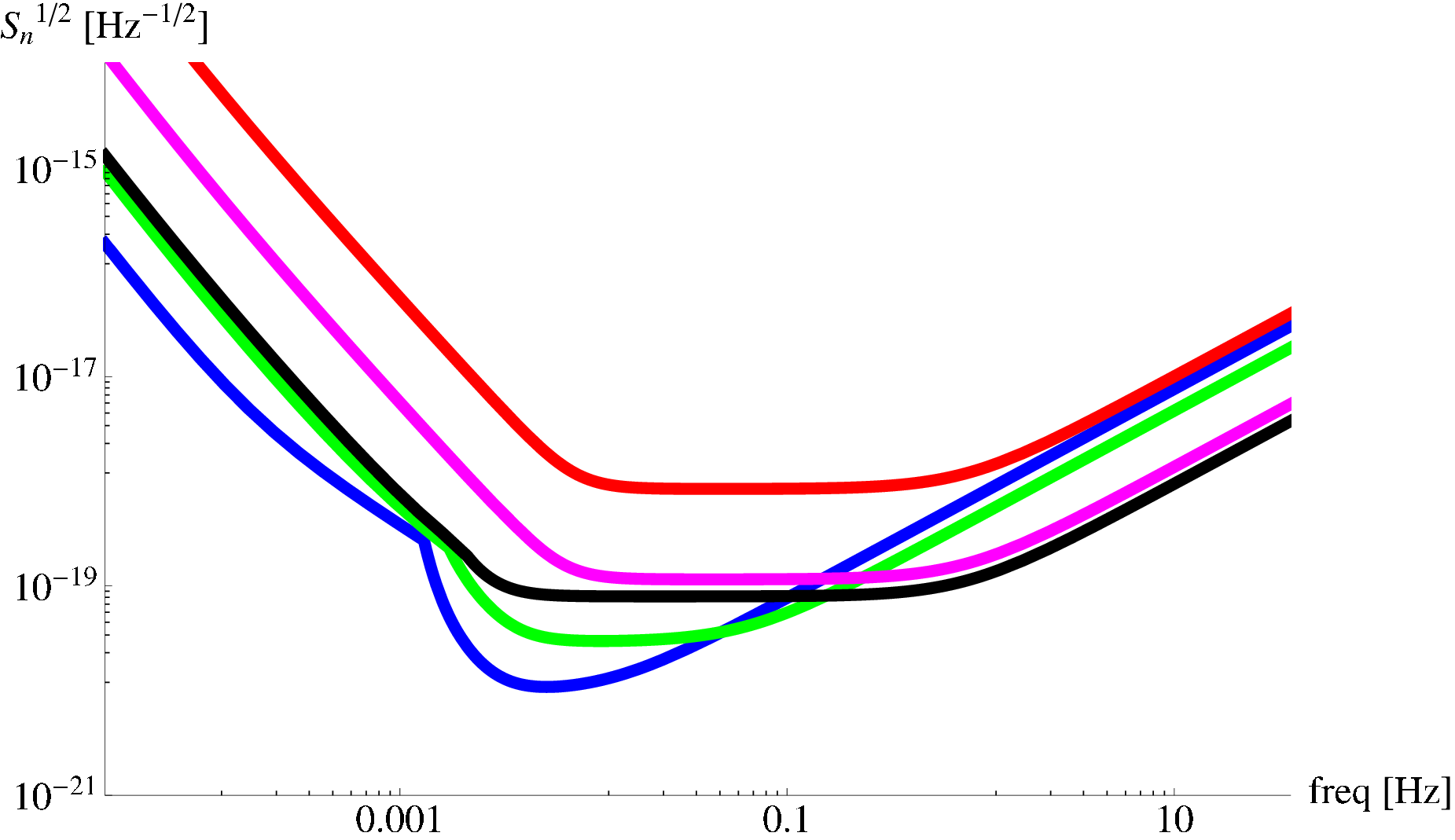}
\caption{Comparison of the rms sensitivity of several mission designs.  The blue curve shows the nominal LISA design
with white dwarf binary confusion noise, the green curve shows the final NGO mission configuration, and the red, magenta, and black
curves show the concept described here, with 10, 1, and 0.1 times the acceleration noise permitted under the DRS specifications
for the nominal LISA design, respectively.
\label{fig:sens}
}
\end{figure}
\end{center}

\section*{\sc Science Performance}

Like LISA, GADFLI will be optimally sensitive to massive black-hole binary inspiral-merger-ringdowns occurring at the hearts of merged galaxies.
In Fig.~2, we plot contours of SNR for LISA and for the three variations of the proposed concept for these
signals.  In addition,
we plot contours of the number of events for two merger-tree models, corresponding to small and large initial mass seeds \cite{Sesana}.
Given the superior sensitivity of GAD-HI compared to NGO, and the comparable sensitivity of GAD-MED (see Fig.~2), we expect both
of these designs to be sensitivity to extreme mass ratio inspirals (EMRIs) and individually resolvable galactic binaries, in addition to comparable
mass massive binaries.  GAD-LO will likely be unable to observe EMRIs or galactic binaries, although it still has a significant
event rate for massive black-hole binary mergers from either merger catalog.

While we do not present detailed parameter estimation at this time, we expect the parameter accuracy to scale roughly inversely with SNR.
As with the nominal LISA design, masses and spins of merging black holes will be measured with extraordinary accuracy,
and any EMRIs that are detected will also provide accurate probes of spacetime, since the complex harmonic structure of
EMRI signals rapidly saturates the parameter measurement capability at a given SNR level.
It is noteworthy that, since the majority of the sky localization accuracy and spin measurement accuracy accumulates at the end of
the coalescence, the greater ease of communication with a geostationary constellation could assist in the earlier identification
of an electromagnetic counterpart to these events.  If a counterpart illuminates during the final merger, a more frequent downlink
of data to the ground, combined with real-time data analysis, could either identify a gravitational wave source earlier, or
localize a known source on the sky earlier.
\begin{center}
\begin{figure}
\includegraphics[width=0.95\textwidth]{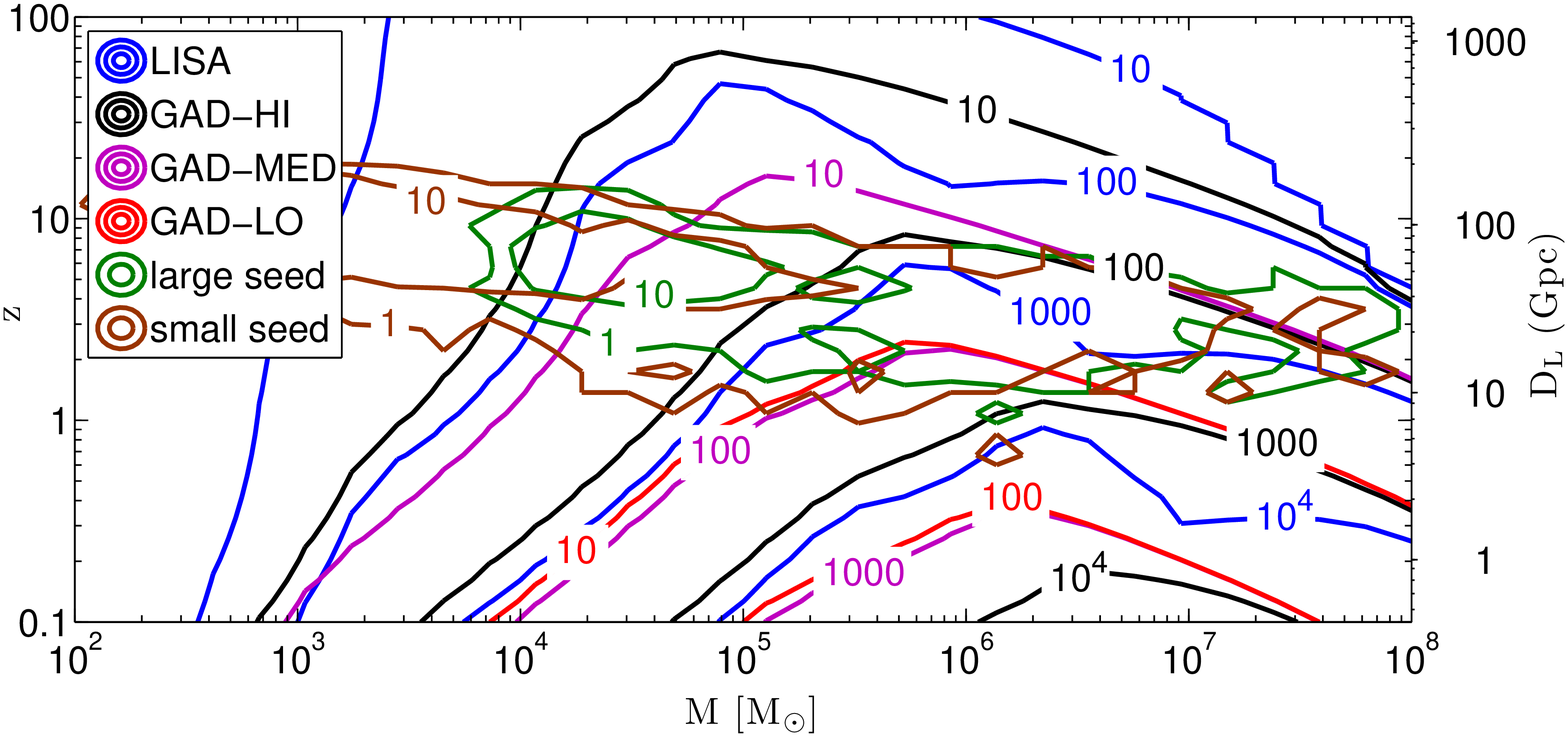}
\caption{Comparison of the signal-to-noise ratios (SNRs) among the nominal LISA design (blue), the $10\times$ DRS GADFLI (GAD-LO, red), the 
$1\times$ DRS GADFLI (GAD-MED, magenta), and the $0.1\times$ DRS GADFLI (GAD-HI, black).  In addition, we plot event number contours
for mergers from merger tree simulations consisting of small seeds (brown, $\langle \log(M)\rangle=3.7$) 
and large seeds (green, $\langle \log(M)\rangle=5.3$).  We note that the contours are all binned in a 30x30 logarithmically-spaced grid, so that
many bins, each containing between 1 and 10 events, lie inside the outermost number contours.  The large seed catalog contains
a total of 720 sources (nearly all of which have SNR$>10$ with GAD-HI), and the small seed catalog contains a total of 2437 sources.
The SNR contours are calculated for an inspiral-merger-ringdown of massive black hole binaries with mass ratios of 3:1,
corresponding to the mean ratio in both catalogs.
\label{fig:hists}
}
\end{figure}
\end{center}

\section*{\sc Costing}

Given our lack of qualifications to perform a proper costing assessment, we will only provide a rough estimate, emphasizing that we expect
the lighter mass and smaller $\Delta$v orbit to be the principle areas of savings relative to the nominal LISA design.  We follow the SGO
costing approach of subtracting cost from the \$1.8B SGO high price point (based on LISA with a less expensive launch vehicle).  We assume
a modest mass savings relative to SGO lowest, so that the Falcon 9 (Block 2) from SpaceX would be the most cost effective
launch vehicle capable of supporting a launch mass slightly below SGO Lowest, for a cost savings of \$300M for savings
on the payload mass and launch vehicle costs.  The 2 year mission lifetime provides a further savings of \$200M in personnel cost.  Like
SGO lowest, GADFLI may not require a propulsion module, and certainly will need far less propellant than OMEGA or heliocentric designs.
We will assume the thrusters available or a less expensive propulsion module will be capable of executing the necessary
120 degree phase change for two sciencecraft that are already on the necessary geostationary orbital trajectories, which would provide
an additional savings of \$100M.  The final cost estimate is therefore \$1.2B, though we again emphasize the rough nature of this estimate,
and the need for a much more thorough assessment by an actual costing expert.

\end{document}

%% file: title.tex
\begin{titlepage}

\begin{center}

\includegraphics[width=0.60\textwidth]{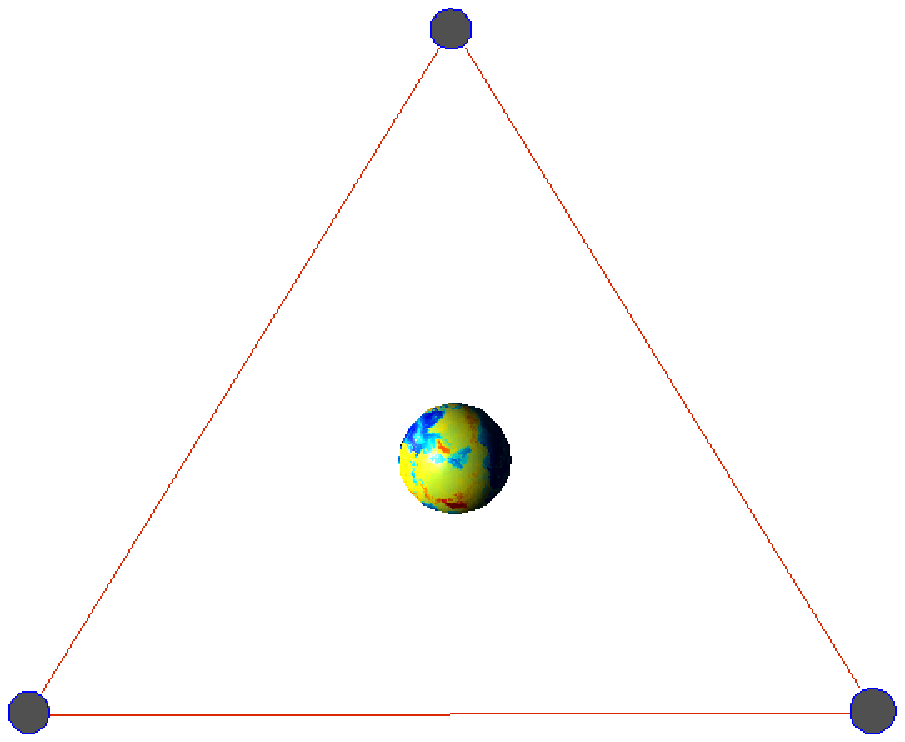}\\[-1.2cm] 

\HRule \\[0.8cm]
\textsc{\huge Geostationary Antenna for}\\[0.1cm]
\textsc{\huge Disturbance-Free Laser}\\[0.1cm]
\textsc{\huge Interferometry (GADFLI)}\\[0.2cm]

\HRule \\[0.6cm]

\textsc{\Large Response to Request for Information:}\\[0.1cm]
\textsc{\Large Concepts for the NASA}\\[0.1cm]
\textsc{\Large Gravitational-Wave Mission}\\[0.5cm]
\textsc{\Large Solicitation Number: NNH11ZDA019L}\\[0.1cm]
\textsc{\Large Category of Response: Mission Concept}\\[0.8cm]

\textsc{Sean T. McWilliams \\ Princeton University}\\[0.3cm]

\includegraphics[width=0.15\textwidth]{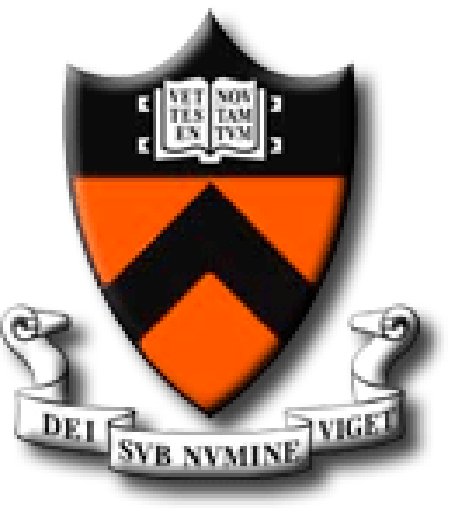}

\vfill

% Bottom of the page
{\large \today}

\end{center}

\newpage

\end{titlepage}